\documentstyle[twoside,fleqn,espcrc2,epsf,psfig]{article}

\newcommand{\AmS}{{\protect\the\textfont2
  A\kern-.1667em\lower.5ex\hbox{M}\kern-.125emS}}

\title{CMS Central Hadron Calorimeter}

\author{Howard S. Budd\address{Department of Physics and Astronomy,
University of Rochester, Rochester, NY 14627}}

\begin{document}

\begin{abstract} We present a description of the CMS central
hadron calorimeter. We describe the production of the 1996
CMS hadron testbeam module. We show the results of the
quality control tests of the testbeam module. We present 
some results of the 1995 CMS hadron testbeam. 
\end{abstract}
\maketitle

\nopagebreak
\section {CMS Central Hadron Calorimeter}

\begin{figure*}
\begin{center}
\vspace{0.in} 
\epsfxsize=5.8in
\mbox{\epsffile[50 120 802 613]{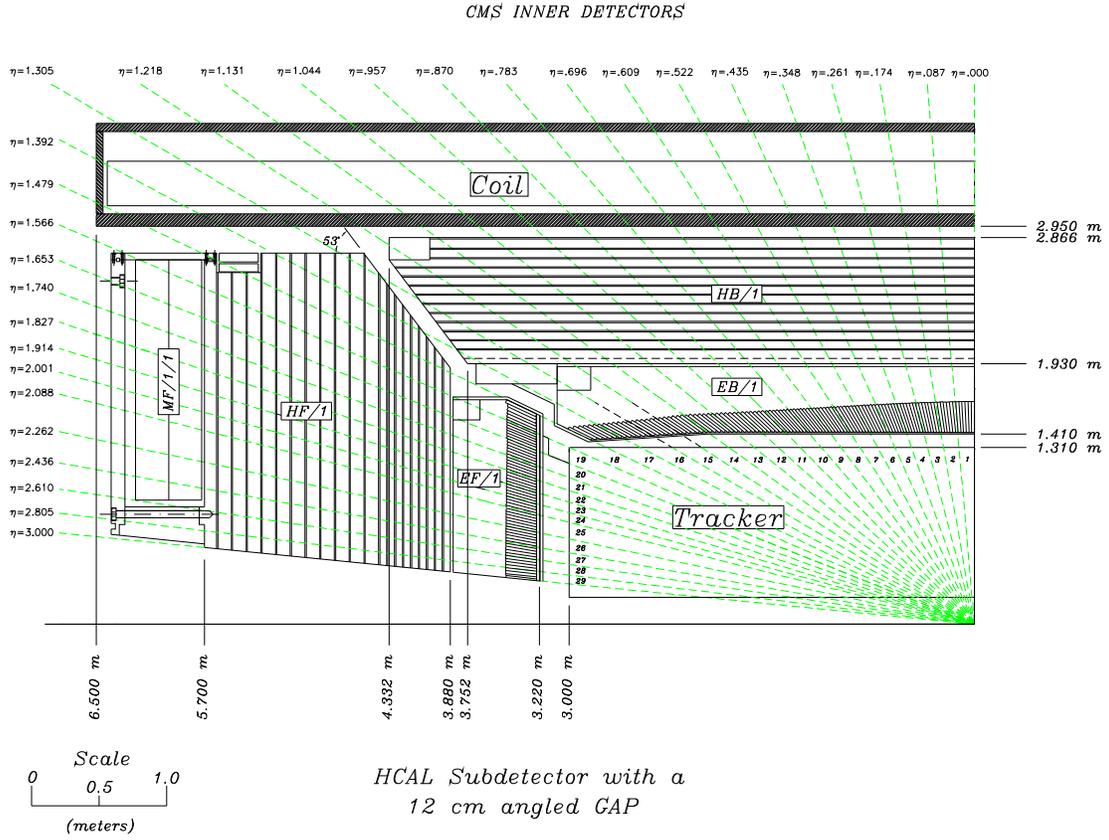}}
\caption{CMS electromagnetic detector, hadronic detector, and 
solenoidal magnet.}
\label{cms_quarter}
\end{center}
\end{figure*}

The design of the CMS detector starts with the 4 T solenoidal magnet
of length 13 m and inner diameter 5.9 m. The magnet determines many
of the features of the CMS calorimeters, since the CMS calorimeter
is located inside  the magnet. Figure~\ref{cms_quarter} shows a
quarter slice of the CMS  up to the coil. The EM calorimeter, ECAL,
consists of lead tungstate, PbW0$_4$, crystals. The hadron
calorimeter, HCAL, surrounds ECAL. The most important requirement of
HCAL is to minimize the non-gaussian tails of the energy resolution
function. Hence, HCAL design maximizes as much interaction length of
material inside the magnetic coil as possible. Copper absorber
satisfies this requirement, as well as being nonmagnetic. In
addition, copper is fairly low Z, so it does not degrade the muon
momentum measurement. Maximizing the amount of absorber before the
magnet requires minimizing the amount of space devoted to active
medium.  The tile/fiber technology is an ideal choice. It consists
of plastic scintillator tiles read out with embedded wave length
shifting (WLS) fibers. This technology was first developed by the
UA1 collaboration and at Protvino~\cite{fiber_tile}. The system is
being used in the  upgrade of the CDF endcap
calorimeter~\cite{cdf_tile}. It enables HCAL to be easily built with
a tower geometry readout. The entire HCAL area can be instrumented
with no uninstrumented cracks.   

HCAL covers the pseudorapidity range $|\eta|<$3.0 with a barrel
(central) and endcap hadron calorimeter. The barrel calorimeter
covers \mbox{0$<|\eta|<$1.4}, while the endcap calorimeter covers
down to $|\eta|$=3.0. The region to $|\eta|$=5.0 is covered by a
very forward calorimeter.  HCAL is segmented into towers of
granularity  $\Delta\eta\times\Delta\phi$=0.087$\times$0.087.  This
granularity is sufficient for good dijet mass resolution.  The rest
of this article  will only discuss the barrel hadron calorimeter.

\begin{figure*} 
\begin{center}
\vspace{0.in}
\epsfxsize=3.4in
\mbox{\epsffile[50 190 760 590]{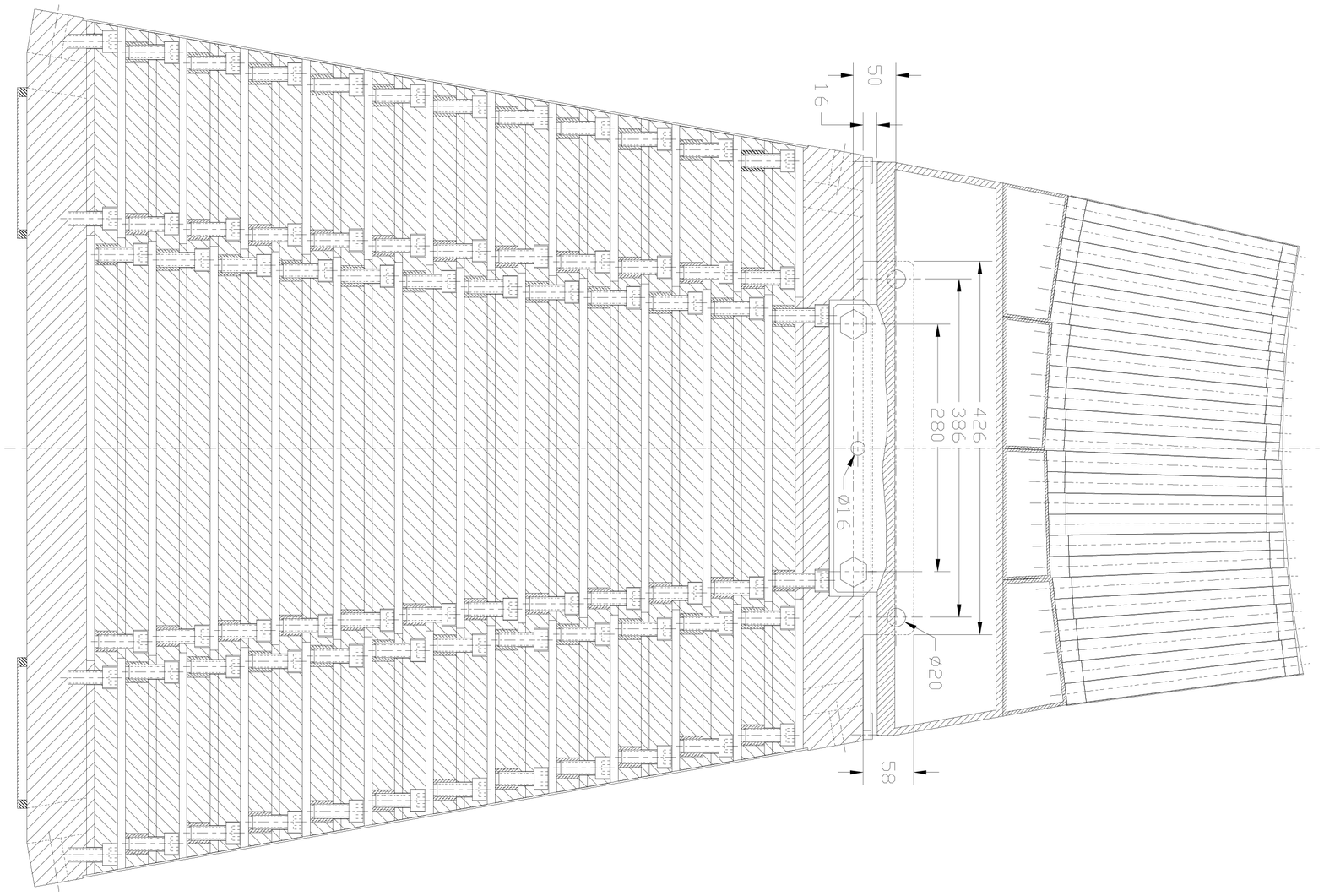}}
\caption{Drawing of 20$^o$ wedge.}
\label{hcal_wedge}
\end{center}
\end{figure*}

\begin{figure*}
\begin{center}
\vspace{0.in}
\epsfxsize=6.5in
\mbox{\epsffile{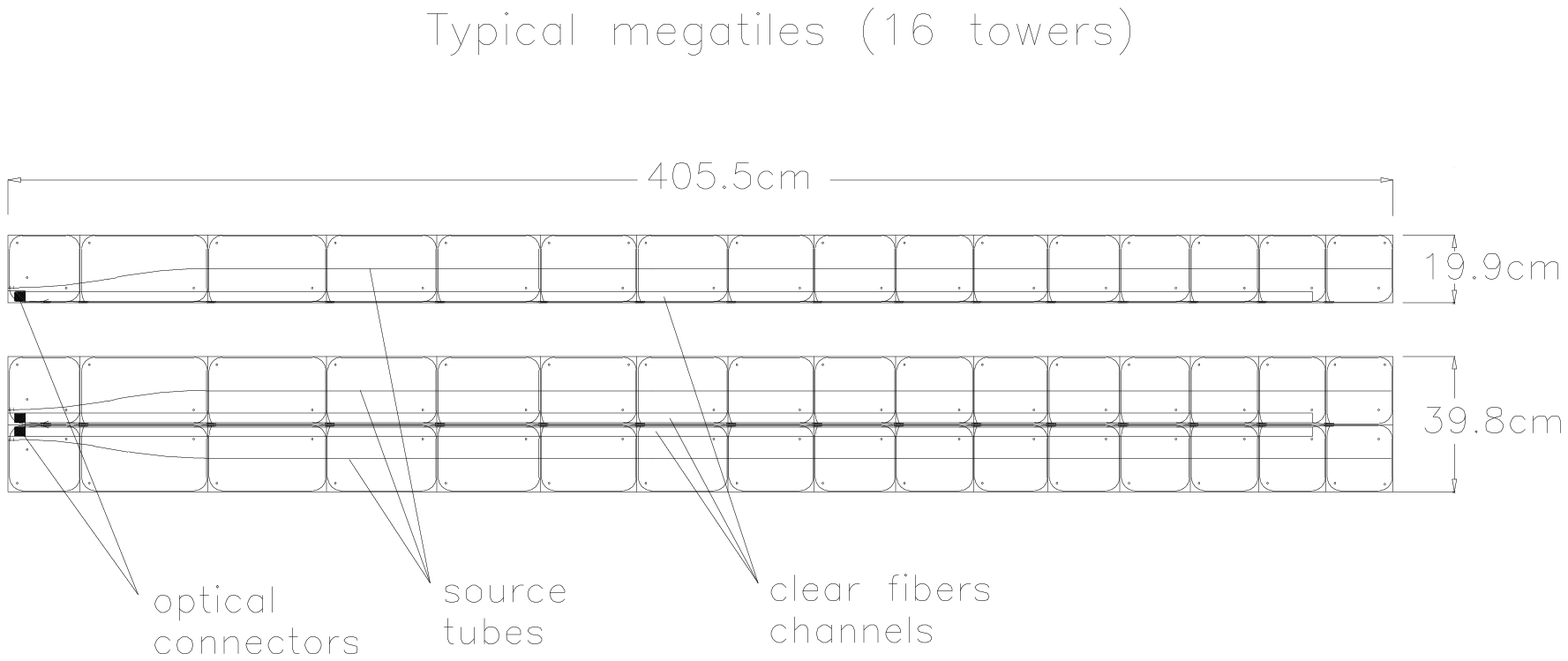}}
\caption{CMS electromagnetic detector, hadronic detector, and
solenoidal magnet.}
\label{tile_tray}
\end{center}
\end{figure*}

HCAL is built as 20$^o$ wedges inside the magnet. 
Figure~\ref{hcal_wedge} shows a 20$^o$ wedge, with ECAL on the  
right side  and HCAL on the left side.  The body of wedges are
copper, but the inner and outer plates are stainless steel for
structural strength. The wedges  are bolted together. This enables
the wedges to be assembled cheaply and minimizes the crack between
the wedges to $\leq$ 2 mm. The wedges contain 15 active layers
segmented into 2 readouts.  A space of 9 mm is left for active
layers. The HCAL   is about 1 meter thick inside the magnet,
6.2~$\lambda$ at $\eta$=0. This is not enough material to contain a
high energy hadron shower.   Hence, 2 layers are instrumented
outside the magnet. A layer of scintillator is put after the coil
and after the innermost  muon steel. This adds an additional 4
$\lambda$ at $\eta$=0.

\begin{figure}
\begin{center}
\epsfxsize=3in
\mbox{\epsffile{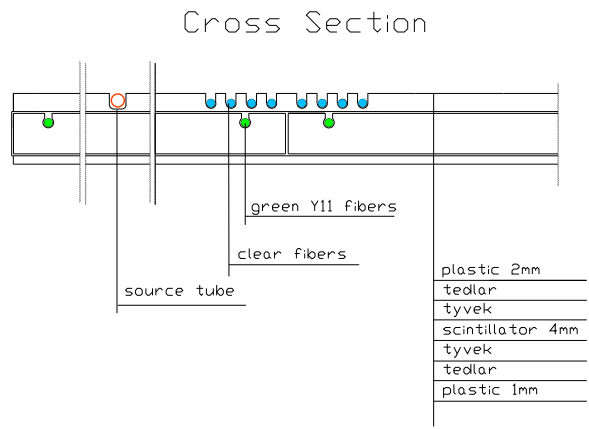}}
\caption{Cross section view of a megatile.}
\label{CROSS_VIEW}
\end{center}
\end{figure}

The active medium uses the  tile/fiber concept as used by the CDF 
plug upgrade. The hadron calorimeter consists of about 70,000 
individual tiles. In order to limit the number of individual
elements, the tiles of a given layer are put into a single 
mechanical unit called a megatile. Figure~\ref{tile_tray} displays a
typical megatile. The megatile with  segmentation of
16($\eta$)$\times$2($\phi$) goes into the $\phi$ center of a wedge
while the megatile with segmentation   16($\eta$)$\times$1($\phi$)
goes into the edge slots in a wedge.  Each layer has 108 megatiles.
Figure~\ref{CROSS_VIEW} shows a cross section of the megatile.  The
scintillator is surrounded by Tyvek 1073D for reflectivity.   The
scintillator and Tyvek are surrounded by tedlar for light tightness.
The top plastic routes the fibers and source tubes.  

The baseline scintillator is 4~mm Kuraray SCSN81. The scintillator
is read out by a WLS fiber, Kuraray multiclad Y-11 (Fluor K27),
embedded in the tile in a $\sigma$ pattern. Outside the scintillator,
the WLS fiber is spliced to Kuraray multiclad clear fiber. Next, the
clear fiber goes to an optical  connector at the end of the pan. An
optical cable takes the  light to an optical descrambler. The
descrambler arranges the fibers into readout towers and brings the
light to a hybrid  photomultiplier tube (HPMT). Since the
descrambler will be operating  in the 4 T magnetic field,
conventional phototubes will not work.

The HPMT consists of a photocathode and a PIN diode separated by 
$\sim$ 1~mm. The photoelectrons from the photocathode are accelerated
to the PIN diode by high voltage, $\sim$ 10,000 volts.  The HPMT has
a gain of 1000 to 2000.  If the electric field is aligned with the
magnetic field, they can operate in a magnetic field of 4 T. This
makes them suitable for CMS HCAL. 

\section{CMS hadron calorimeter test beam}

We have built modules for a 1995 and 1996 testbeam at CERN. The
purpose of the testbeam is to determine the response  of the
calorimeter in a magnetic field and to determine  the resolution of
the combined ECAL and HCAL calorimeter.   To determine the response
in a magnetic field, the calorimeter was put  inside the EPS magnet
in the H2 beam line at CERN. We describe the calorimeter and its
production to illustrate the production of the real device and show
some of the quality control tests we will use for the real device.
All numbers we quote for the quality control are for the 1996
testbeam module.

\begin{figure} 
\begin{center}
\vspace{0.in}
\epsfxsize=2.8in
\mbox{\epsffile[86 70 452 451]{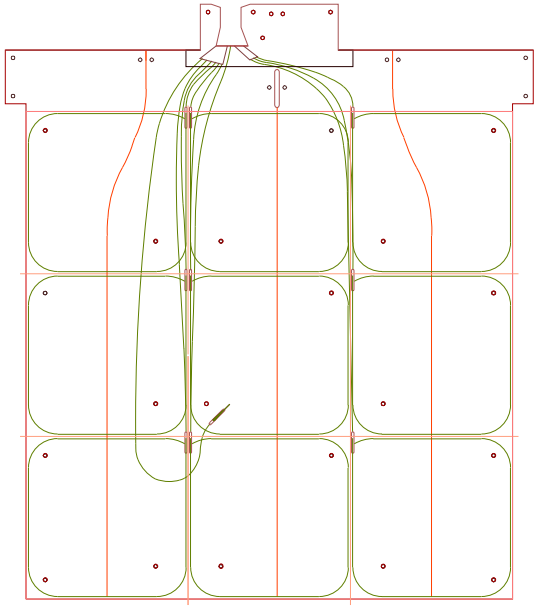}}
\caption{ Megatile for 1996 CMS HCAL testbeam.}
\label{tile_96}
\end{center}
\end{figure}

\subsection{1996 HCAL Testbeam Megatile} 
Figure~\ref{tile_96} shows the megatile for the 1996 testbeam. The
active area is 64 cm by 64 cm and consists of 9 separate towers. The 
transverse size of the testbeam megatile is determined by size of 
the center of the EPS magnet.  Embedded in each tile is a WLS fiber
in the $\sigma$ pattern. Outside the tile, the WLS fibers are
spliced to multiclad clear fibers. The clear fibers are  routed out
to a connector at the end of the megatile.  The fiber on the left
going into the middle tile is a light injection fiber. For
calibration purposes light is injected into the tile through this
fiber. The megatile has 3 thin stainless steel tubes, diameter=1 mm,
that will route Cs$^{137}$ radioactive sources through the center of
each tile. A picoammeter measures the current generated by the
source to simultaneously calibrate the tile, fibers, and
photodetector. 
   
The elements for the megatile are drawn on a CAD system. The drawings
are converted to machine code which can be run on a computer controlled
router, a Thermwood machine. 

We describe the production and quality control of the device. First
the WLS fibers are cut, polished, and mirrored. The reflectivity of
the mirror is  checked by measuring test fibers which are mirrored
along with the fibers used in the calorimeter. Measuring the
reflectivity of the mirror gives  
$$light_{with~mirror}/light_{with~mirror~cut~off} = 1.85$$ 
\noindent 
This measurement is done with a computer controlled UV scanner  with
the fibers read out by pin diodes. Clear fibers are spliced onto WLS
fibers with a fusion splicer. The transmission across the splice is
checked by splicing a sample of WLS fibers onto WLS fibers. The
splice region is measured with the UV scanner. The transmission
across the splice is  92.6\% with an RMS of 1.8\%. Next, the optical
fibers are glued into a 10 fiber connector. This configuration is
called a pigtail. In order to get the fiber lengths correct, the
pigtail is assembled in a templet. The connector is diamond
polished. The fibers are  measured with the UV scanner. The scanner
checks the  green fiber, clear fiber, splice, and mirror. The RMS of
the light  from the fibers is 1.9\%. 

The pigtail is inserted into the megatile. The completed megatile is
checked with an automated source scanner. A Cs$^{137}$ source is in
a lead collimator. This yields a 4 cm diameter source spot on  the
megatile. The collimator is moved with a computer controlled x-y
motor. From the scanner we determine the relative light yield  of
each tile and the uniformity of the each megatile. The gain of the
individual tiles has an  RMS of 4.6\%, while the transverse
uniformity of the megatile is 4.5\%.  A Cs$^{137}$ wire source is
run through the 3 source tubes and the light yield is measured. The
RMS of the ratio of collimated source  to wire source is 1.3\%. This
means the line sources, which can  be used when the calorimeter is
completely assembled, can calibrate the tiles to better than 1.3\%.
Fiber cables are constructed  separately and tested with the UV
scanner. 

The testbeam module has several methods to maintain the calibration.
As just described, it has small steel tubes to run a wire source. 
It has a laser and LED light injection system. The photomultiplier
tube that reads out the calorimeter has an addition fiber going to
it. Either  a laser or an LED can inject light into this fiber.
During the testbeam  we injected light with both the laser and LED.
As shown on the  drawing of the megatile, there is a fiber which can
inject light into a tile. A more complete description of the
calibration, including procedures for calibrating the full
calorimeter,  is given in ref~\cite{freeman}.

\subsection{1995 HCAL Testbeam Results} 
\begin{figure}
\begin{center}
\epsfxsize=3in
\mbox{\epsffile[24 37 541 522]{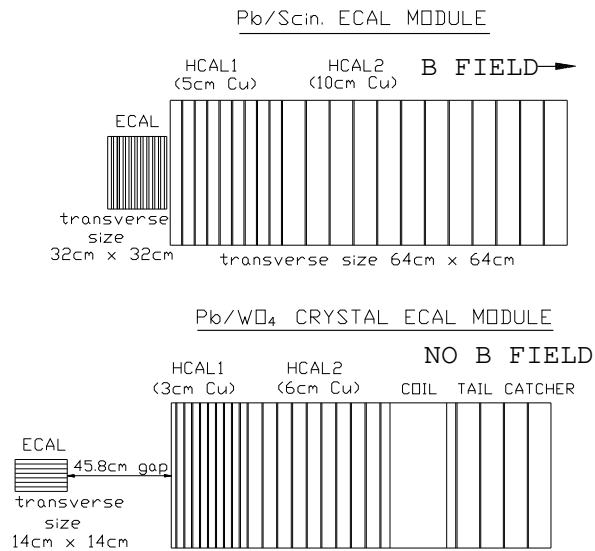}}
\caption{Schematic drawing of the  calorimeter modules used
during the 1995 test beam.}
\label{testb5}
\end{center}
\end{figure}

We present results from the 1995 testbeam. Results for the 1996
testbeam are not yet final. Figure~\ref{testb5}   shows the testbeam
configurations. For the Pb/scin ECAL module,  ECAL is the  10 layer
lead/scintillator sandwich sampling calorimeter, positioned directly
upstream of the HCAL. The HCAL corresponds to the end cap
configuration (5~cm and 10~cm copper absorber plates in HAC1 and
HAC2 respectively). For the PbWO$_4$ crystal  ECAL module, the ECAL
detector (7 $\times$ 7 matrix of 2~cm by 2~cm by 23~cm PbWO$_4$
crystals) is followed by the HCAL in the barrel configuration (3~cm
and 5~cm copper absorber plates in HAC1 and HAC2 respectively).
There is approximately 46~cm air gap between the end of ECAL and the
front face of HCAL.  Each layer for both configurations is readout
by a phototube connected to a 10 m optical cable.

\begin{figure}
\begin{center}
\epsfxsize=3in
\mbox{\epsffile{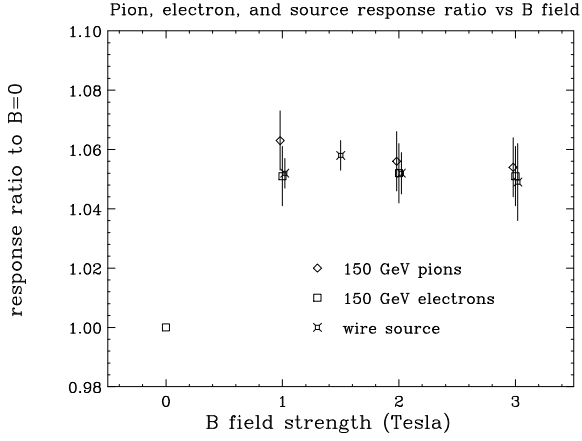}}
\caption{H2 Test Beam results:
average pion, average electron, and Cs$^{137}$ 
source response of the calorimeter, as a function of the B field.}
\label{h2_bfield}
\end{center}
\end{figure}

Figure~\ref{h2_bfield} shows the average pion and electron response
of the calorimeter, as a function of the B field, relative to the 0
T setting. The data indicates an approximate 6\% increase in the
response of the calorimeter to pions and electrons for B field at
1~T, 2~T, and 3~T. The measurement is consistent with the increased
light yield of the scintillator tiles, as measured by the
calibration system using the radioactive $\gamma$ source. The 1996
test beam run had the magnetic field transverse to the beam
direction. The result is different if the magnetic field is
transverse to the  beam direction. A discussion is in 
ref~\cite{freeman} and will be in future presentations of the 1996
test beam.

The overall calibration constant of the hadron
calorimeter is determined by using 50 GeV hadrons, and the overall 
calibration constant of EM calorimeter is determined using 
150 GeV electrons. The total energy of pions is defined as 
\[
E_{tot}^{pions} = \alpha\times E_{em} + E_{had}
\]
                  
\begin{figure}
\begin{center}
\epsfxsize=3in
\mbox{\epsffile{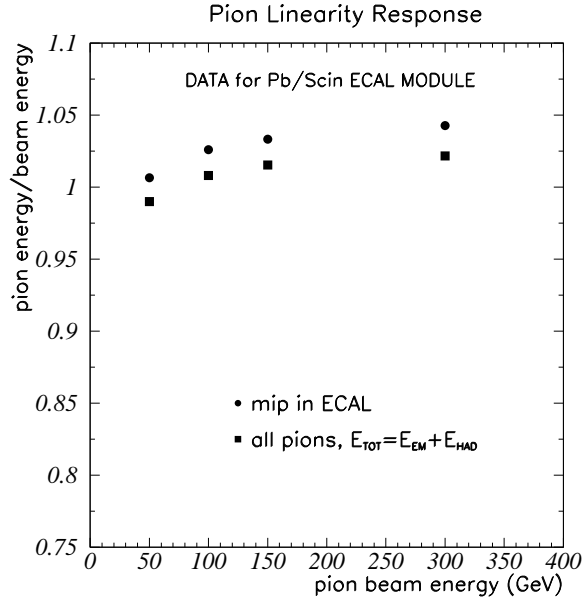}}
\caption{Linearity of Pb/scin ECAL module.}
\label{h2_linearity}
\end{center}
\end{figure}

\noindent 
where $\alpha$ is a parameter we vary to study linearity and resolution.
Figure~\ref{h2_linearity} shows the linearity 
of the Pb/scin ECAL configuration 
for pions which are minimum ionizing (mip) in  ECAL and  all pions ($\alpha$=1).
The energy resolutions for the 2 cases are comparable.
The fit of this data to the relative energy resolution function is
\[
        \sigma_{E}/E = (stoch.~term)/\sqrt{E} \oplus (const.~term)
\]
where stochastic and constant terms are combined in quadrature.
For pions which are mip in ECAL we have
\[
        \sigma_{E}^{\pi~mip~in~ECAL}/E = 
\]
\vspace{-.24in}
\[      
        ~~~~~~~~~~~~(90 \pm 0.1) \%/\sqrt{E} \oplus (4.8 \pm 0.1)\%.
\]
For the case of all pions we have
\[
        \sigma_{E}^{all~\pi}/E = 
\]
\vspace{-.24in}
\[
        ~~~~~~~~~~~~(77 \pm 0.1) \%/\sqrt{E} \oplus (5.5 \pm 0.1)\%.
\]
At low energies the energy resolution  of all pions is narrower than
pions which are mip in ECAL due to the fine sampling of
the ECAL calorimeter.

\begin{figure}
\begin{center}
\epsfxsize=2.8in
\mbox{\epsffile{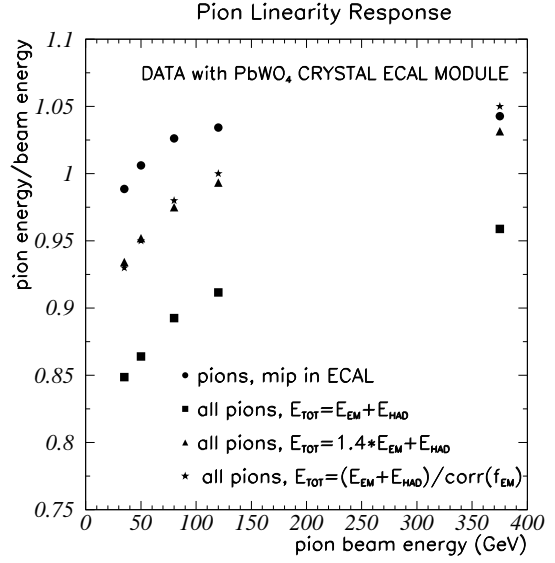}}
\caption{Linearity of PbWO$_4$ crystal module for hadrons.}
\label{h4_linearity}
\end{center}
\end{figure}

\begin{figure}
\begin{center}
\epsfxsize=2.7in
\mbox{\epsffile{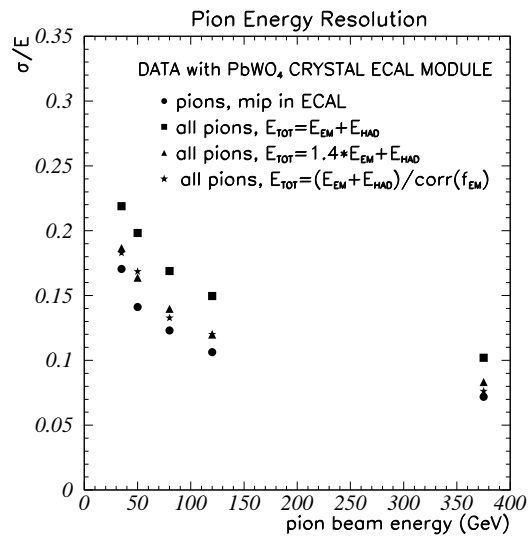}}
\caption{Resolution of of PbWO$_4$ crystal ECAL module for hadrons.}
\label{h4_res}
\end{center}
\end{figure}

Figure~\ref{h4_linearity} and Figure~\ref{h4_res} show the 
linearity and resolution of the  PbWO$_4$ crystal configuration  for
hadrons mip in ECAL and all hadrons. Hadrons mip in ECAL have the
same linearity and  resolution in both configurations.  However, 
the PbWO$_4$ crystal data for all pions ($\alpha$=1) is non-linear.
The average response of the combined ECAL+HCAL calorimeter to pions
is also approximately 10\% lower than the HCAL only calorimeter. The
resolution is much worse for all pions than pions mip in ECAL.  The
linearity and resolution is shown for $\alpha$=1.4. Although both
the linearity and resolution are improved, this type of energy sum
cannot be used for jets.

\begin{figure}
\begin{center}
\epsfxsize=2.7in
\mbox{\epsffile{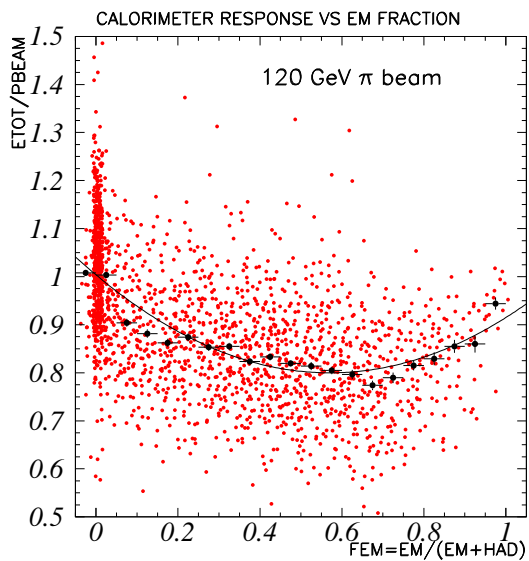}}
\caption{Energy response to pions,
as a function of ECAL fraction for the PbWO$_4$ crystal ECAL module.}
\label{fem_120}
\end{center}
\end{figure}

Figure~\ref{fem_120} shows the scatter plot of the total energy
response of the calorimeter normalized the beam momentum,
$E_{TOT}/p_{beam} =(E_{EM} + E_{HAD})/p_{beam}$ as a function of the
fraction of energy deposited in the ECAL, $f_{EM} = E_{EM}/(E_{EM} +
E_{HAD})$. Data points with $f_{EM} \approx$ 0 correspond to pions
that interacted in the HCAL. Data with $f_{EM} \approx$ 1
correspond to pions fully contained in the ECAL. A fit of the data
binned as a function of $f_{EM}$ to a second degree polynomial is
shown on the figure. The fitted function $corr(f_{EM})$ is used
to correct the total energy.  
\[ 
E_{TOT}^{cor} = (E_{EM} +E_{HAD})/corr(f_{EM}) 
\] 
The linearity and resolution are shown on
Figure~\ref{h4_linearity} and~\ref{h4_res} for the $f_{EM}$ corrected
data. The correction improves the linearity and resolution. 

\samepage
\nopagebreak

\vspace{-1in}

\end{document}